\documentclass[superscriptaddress,showpacs,preprintnumbers,amsmath,amssymb,floatfix,elsart,prd]{revtex4-1}

\usepackage{graphicx,subfigure,color,hyperref}
\usepackage{dcolumn}
\usepackage{bm}
\usepackage{epstopdf}

\newcommand{\beq}{\begin{equation}}
\newcommand{\eeq}{\end{equation}}
\newcommand{\bey}{\begin{eqnarray}}
\newcommand{\eey}{\end{eqnarray}}

\begin{document}

\title{Thin accretion disks around traversable wormholes}

\author{Farook Rahaman}
\email{rahaman@iucaa.ernet.in} \affiliation{Department of
Mathematics, Jadavpur University, Kolkata 700032, West Bengal,
India}

\author{Tuhina Manna}
\email{tuhina.manna@sxccal.edu} \affiliation{Department of Mathematics, St.Xavier's College(Autonomous), 30 Mother Teresa Sarani, Kolkata-700016,
India}
\author{Rajibul Shaikh}
\email{	rshaikh@iitk.ac.in} \affiliation{Department of
Physics, IIT, Kanpur,
UP}

\author{Somi Aktar}
\email{somiaktar9@gmail.com} \affiliation{Department of
Mathematics, Jadavpur University, Kolkata 700032, West Bengal,
India}

\author{Monimala Mondal}
\email{monimala.mondal88@gmail.com} \affiliation{Department of
Mathematics, Jadavpur University, Kolkata 700032, West Bengal,
India}

\author{Bidisha Samanta}
\email{samantabidisha@gmail.com} \affiliation{Department of
Mathematics, Jadavpur University, Kolkata 700032, West Bengal,
India}

\date{\today}

\begin{abstract}
In this paper, we aim to investigate various physical properties and characteristics of radiation emerging from the surface of accretion disks, in a rotating traversable axially symmetric wormhole spacetime of the Teo class. We have studied the marginally stable orbits and accretion efficiency graphically, corresponding to different values of dimensionless spin parameter $J/M^2$ ranging from 0.2 to 1.5 and some values of the throat radius $r_0$, in comparison to the Kerr black hole with the  same parameter values, and also tabulated the results. The energy flux radiated by the accretion disk $F(r)$, the temperature distribution $T(r)$ and the emission spectra $\nu L(\nu)$ is plotted, corresponding to varying values of the dimensionless spin parameter $J/M^2$ and throat radius $r_0$. Also, the critical frequency at which the luminosity attains its maximum value, for various values of the angular momentum of the wormhole $J/M^2$ and $r_0$ is tabulated. Lastly,  we have employed ray-tracing technique, to produce the intensity map of the image of an accretion disk, as observed by an asymptotic observer, under two conditions: firstly when the disk is on the same side as the observer and we have also compared those with the images of an accretion disk in case of Kerr black hole with same parameters. Secondly, the images have been provided when the disk and observer are on opposite sides of the throat. This study may help to detect and distinguish wormhole geometries from other compact objects.
\end{abstract}


\maketitle

\section{Introduction:}
It is argued that the mass of almost all  astrophysical bodies  develop through accretion.  The current observation of the center of the galaxy M87  through the  Event Horizon Telescope  suggests  that in most of the active galactic nuclei (AGN's), there exist gas clouds around the galactic center, together with an accompanying accretion disc. These gas clouds (scales from a tenth of a parsec to a few hundred parsecs )   are expected to develop  a geometrically and optically distorted disc that  engrosses maximum of the ultraviolet radiation and the soft X-rays.   Significant  astrophysical evidence can be obtained  from the observation of the motion of the gas flows in the gravitational field of super massive objects.  The temperature in the innermost region of the disk is about  $10^7  K $  and from this hot region,  the X-ray emission arises. To study the  strong field gravity, one can use these highly compact objects,  as
the detected emission arises from regions close to these  compact objects. According to the standard theory, the accretion disk is presumed  to be an optically thick Newtonian type. This model suggests that  the local emergent flux ( just like  a black body ) is  associated with the energy decadence
at  a  specific  radial  point  in  the  disk. The  sum of black body components coming  from various
radial  positions  in  the  disk yields the  observed spectrum.  Usually,  general relativistic issues change this Newtonian spectrum by incorporating Doppler broadening,  gravitational  redshifts,  and  light  bending.\par
Since wormholes \cite{kg1}, \cite{kg14} are  a specific type of astrophysical objects, so it has been an active area of research to study accretion phenomenon around wormholes.  The phenomenon of accretion around compact objects is an outstanding example for studying the signatures of wormholes versus black holes.  Initial research on accretion disks using a Newtonian approach was  conducted  by Shakura and R.A. Sunyaev.\cite{SS}  Later Thorne et. al. \cite{first} developed  a general relativistic model of the  thin accretion disk, where the accreting matter   had  a Keplerian motion. They assumed the disk is in a steady-state, i.e., the mass accretion rate $\dot{M}$ does not change with time, as well as  it is   independent of the radius of the disk. From   a  theoretical point of view, the efficient cooling due to radiation transport,    maintains  this disk in  hydrodynamic and thermodynamic equilibrium, thus ensuring a black body electromagnetic spectrum.  The spectra  from  the  sources  may provide some  information   on  the geometry and dynamics of innermost regions of accretion disks.  Following Page-Thorne \cite{Page} model, one can study  emissivity properties such as the luminosity spectra,
flux of radiation, temperature profile  using  accretion around wormholes \cite{Luminet}. Further radiation properties of thin accretion disks were analysed in \cite{Thorne, Gair}, along with the effects of the photon capture by the black hole on the spin evolution. The emissivity properties of the accretion disks, which helps us to study the conversion of rest mass into emitting radiation in the accretion disk  were  studied in \cite{Bhat}, \cite{Torres}, \cite{Yuan}, \cite{Guzman}. The work in these papers  have  helped to distinguish between the accretion properties around different classes of exotic central objects, such as non-rotating or rotating quark, boson or fermion star, by studying the radiation power per unit area, the temperature of the disk and the emission spectrum of the
emitted radiation and comparing them with the Schwarzschild and Kerr black hole. Through the investigation of differences in the thermodynamic and electromagnetic properties of the accretion
disks, Kovacs et. al.  \cite{K}, identified the observation signatures that may help to distinguish between black holes and naked singularities. Image of the Janis-Newman-Winicour naked singularity with a thin accretion disk was studied in \cite{Gyul}. They concluded that for some specific values of the spin parameter and of the scalar charge, the radiation energy flux and conversion efficiency from the disk around a rotating naked singularity is higher than that around a Kerr black hole. Accretion around rotating wormholes is an active area of current research. Recently, gravitational wave echoes of the Damour-Solodukhin wormhole (RDSWH)  have been  studied by Bueno et al. \cite{Bueno}. Nandi et.al. further studied
a rotating generalization of the RDSWH\cite{Nandi} and concluded that RDSWH are experimentally indistinguishable from Kerr black hole by accretion characteristics.
Similar works on comparative study of thin accretion disks around rotating gravastars and Kerr-type black holes, in order to distinguish between them, was undertaken by\cite{Harko2}. It was concluded that gravastars provide a less efficient mechanism for converting mass to radiation than black holes. Many major works in modified theories of gravity are also available in literature. The physical properties of a thin accretion disk in the static and spherically symmetric spacetime metric of vacuum $f(R)$ gravity was analysed  in  \cite{Pun}. Thus astrophysical observations of the emission spectra from accretion disks may lead us to the possibility of directly
testing modified gravity models. Several classes of static and rotating brane
world black holes  have been  investigated in \cite{Pun2}, and compared with the general
relativistic case. The thin accretion disk features in dynamical Chern-Simons modified
gravity  have been  studied extensively in \cite{Harko3} and in Horava-Lifshitz gravity in \cite{Harko4}.   An interesting work    on thin accretion disk around a rotating Kerr-like black hole in Einstein-bumblebee gravity model has been done by   \cite{Bumblebee}.  Research on the properties of the radiation emerging from the surface of the accretion disk around static and rotating axially symmetric wormhole geometries were conducted in \cite{Harko5}, \cite{Harko6}. In \cite{Paul} numerically constructed images of thin accretion disks in rotating wormhole backgrounds, for the Kerr-like and the Teo class of wormholes was studied, and interestingly it showed that there are dramatic  differences between accretion disk images in wormhole backgrounds, compared to black hole ones, owing to the fact that a wormhole can theoretically have accretion disks on both sides of its throat.    Although, \cite{Paul}  gives a full explanation of using ray-tracing of a general Teo wormhole, they have considered specific forms of the metric functions in order to generate results and images. In our present work, we have not merely changed parameters, but have taken completely different forms of the metric functions, thereby giving a wormhole with completely different spacetime geometry or gravitational field within the Teo class. Therefore, within the Teo class, depending on the forms of the metric functions, we could have various wormholes with various spacetime geometries or gravitational field, provided the forms of the metric functions satisfy the conditions obtained by Teo. The observational signatures in these various cases in principle will be different. The specific example which we have considered is completely different from the one considered in \cite{Paul}. Moreover, the focus of \cite{Paul} was on constructing the images, whereas our focus was not only on the images but also on the other properties such as luminosity, critical frequency etc. We believe that the results presented in the paper will indeed be relevant in future observations for the reasons explained there.

  The motivation of the study is as follows.  It is well-known that, according to the Kerr Black Hole Hypothesis, the spacetime geometry around astrophysical
compact objects is well described by the Kerr solution of General Relativity (GR). However, in modified
gravity theories or in presence of quantum effect or exotic matter, one can expect deviation from the
Kerr spacetime. Such deviations should be tested using observations. Therefore, it is astrophysically relevant to study various observational aspects of other possible geometries beside those of the Kerr solution. The motivation behind such studies is to see whether to what extent we can distinguish between a Kerr black hole and a deviated non-Kerr spacetime. In cases where differences in the observational signatures arise, we can distinguish a Kerr black hole from a non-Kerr geometry. However, in cases where differences do not arise, a non-Ker geometry mimics a Kerr black hole. In our work, we have studied a similar aspect by considering a wormhole and a Kerr black hole. Although  in the literature, there are several wormhole spacetimes which are obtained as exact solutions of GR as well as other modified gravity theories, and matter for such wormholes are known, most of these wormholes are non-rotating solutions. This is due to difficulties associated in obtaining rotating solutions as exact solutions. Given this difficulty, researchers often consider rotating wormhole models which are not exact solutions of any gravity theory, but can be used in modelling horizonless compact objects. The motivation behind this latter approach is comparative studies between a black hole and a wormhole through their observational aspects. In other words, the objective in this case is not the matter on the RHS of the field equations, but to see whether and to what extent we could distinguish between a wormhole and a black hole or to test the Kerr Black Hole Hypothesis through observational aspects. The literature is replete with several such kind of approaches. In our work also, the motivation was to explore comparative studies between a wormhole and a black hole through observational signatures of accretion disks around them.

\par
The paper is organized as follows: in section II we have described some physical properties of thin accretion disks, followed by the investigation of marginally stable circular orbits in section III. Next in section IV, we have studied the accretion disk properties of a rotating traversable wormhole of the Teo class in details. Following this, in section V, the images of an accretion disk around a wormhole is analysed as observed by an asymptotic observer whose location is taken first on the same side of the throat as the disk; and again on the side of the throat opposite to that of the disk. Lastly, in section VI we have mentioned some concluding remarks.

\section{ Some physical properties of thin accretion disks}

Usually, it is claimed that compact X-ray sources are binary systems. One is a normal star and the other is a highly compact object     and former  landfills material onto its  latter  companion. The majority of the standard  models for this  mass transfer   suggest  that the relocated  material forms a thin disk around the highly compact objects. The  viscous stresses handover angular momentum  outside over  the disk, causing    the material to spiral steadily inward. Also, these  viscous stresses, functioning  against the disk's differential rotation,    warm the disk thus    producing    a large flux of X-rays. According to   steady-state models,  the disk accretion ( in which accreting material is interstellar gas and magnetic fields )   onto a highly compact object may  happen in the nuclei of  galaxies. The gravitational energy  of the   accreting material   converts to heat and  some parts of the
heat transform into
radiation. This radiation   partly seeps out  and chills  the
accretion disc.  After receiving  this  radiation  and analyzing  its electromagnetic spectrum through  radio, optical and X-ray telescopes ,  we get    information about the disk as well as compact object surrounded by the disk. In the analysis  of mass accretion, we adopt a geometrically  thin  accretion  disk. In  the thin accretion disk, the vertical size of the  disk is insignificant   compared to  its horizontal extension i.e. most of the matter lies close to the radial plane.  This means that  the disk height  $H$ ( described  by the maximum half-thickness of the disk ),  is considerably  lesser than any characteristic radii $R$ of the disk,  $\frac{H}{R}  << 1$. Here, the vertical size in cylindrical coordinates $(r, \phi, z)$   defined along the z-axis
is negligible,    Also for this consideration,  stress and dynamic friction in the disk  generates heat  that can be spread through the radiation above  its surface.   This cooling process   confirms that the disk could  be in hydrodynamical equilibrium and as a result,   the mass accretion rate $\dot{M}$  in the disk does not change with time i.e. constant.  Since the disk is   steadied at hydrodynamic equilibrium,  the pressure and vertical entropy gradient should be  negligible.  This  thin disk has an inner edge defined by the marginally stable circular
radius  $r_{ms}$  and the accreting matter at higher radii has a Keplerian motion. This thin accretion  disk supposition implies that spacetime metric components  as well as the accretion particles move  in Keplerian orbits around the compact objects   with a rotational velocity,
$ \Omega  = \frac{d \phi}{dt} $   and its
  specific energy   E  $\&$ specific angular momentum  L  depends only    on the radii of the orbits i.e. on the radial coordinate.
The accretion particles with the four-velocity $ u_\mu$   form a disk and the disk has an averaged surface density $\Sigma$,
\begin{equation}
\Sigma = \int_{-H}^H <\rho_0> dz,
\end{equation}
where $\rho_0$ is the density of rest mass and   H is the disk height during the time $\Delta t$. Here, $<\rho_0>$	 is  defined as the averaged rest mass density  in time $\Delta t$  with  angle $2 \pi $.
The orbiting particles in the disk is modeled by an
anisotropic fluid source which is given by,
\begin{equation}
 T^{\mu \nu } = \rho_0 u^\mu u^\nu + 2 u^{(\mu} q^{\nu)} + t^{\mu \nu },
\end{equation}
 where   the energy flow vector $q_\mu$ and the stress tensor $ t^{\mu \nu }  $
are measured in the averaged rest-frame with $  u_\mu q^\mu  =  0, ~ u_\mu t^{\mu \nu }  =  0. $     Here we neglect specific heat.
Another characteristic of   the disk structure is  the torque which is given by,
\begin{equation}
W_\phi^r = \int_{-H}^H <t_\phi^r> dz,
\end{equation}
with the averaged component $<t_\phi^r>$ in time $\Delta t$  with  angle $2 \pi $.
The most significant physical measure recounting the disk is the radiation flux $ F(r) $ over the disk surface that can be obtained  from the  time and orbital average of the energy flux vector as,
\begin{equation}
F( r )  =   <q^z> .
\end{equation}
Using stress energy tensor     of the disk $ T^{\mu \nu } $, one can define the four-vectors of the energy and of the angular momentum flux  in a local basis  as,
\begin{equation}
 -E^\mu = T^\mu_ \nu  \left( \frac{ \partial }{ \partial t} \right)^\nu    ~and ~ J^\mu = T^\mu_ \nu  \left( \frac{ \partial }{ \partial  \phi} \right)^\nu    ,
\end{equation}
 respectively.

Now, the structure equations of the thin disk could be obtained from  the conservation laws of the rest mass, of the energy, and of the angular momentum.  Rest mass conservation,
\begin{equation}
\nabla_\mu (\rho_0u^\mu) =0
\end{equation}
 yields  the time averaged rate of the rest mass accretion,
which  is independent of the disk radius,
\begin{equation}
\dot{M} \equiv - 2\pi \sqrt{-g} \Sigma u^r = ~\text{constant},
\end{equation}
where   dot implies  the differentiation with respect to the time  coordinate  t.  The energy conservation law,  $\nabla_\mu E^\mu =0 $   yields,
\begin{equation}
\dot{M} E - 2 \pi \sqrt{-g} \Omega W_\phi^r ]_{,r} = 4 \pi \sqrt{-g} F E,
\end{equation}
where a comma represents  the derivative with respect to the radial coordinate r. This is an equilibrium equation, which suggests  that the energy conveyed by the rest mass flow, $\dot{M} E$, and the energy conveyed by the torque in the disk, $2 \pi \sqrt{-g} \Omega W_\phi^r$, is  well-adjusted by the energy emitted   from the surface of the disk, $4 \pi \sqrt{-g}F E$.
Angular momentum conservation law,  $\nabla_\mu J^\mu =0 $, yields,
\begin{equation}
[\dot{M} L - 2 \pi \sqrt{-g}   W_\phi^r ]_{,r} = 4 \pi \sqrt{-g} F L.
\end{equation}
As before, the first term characterizes the angular  momentum supported by the rest mass of the disk; the second term is the angular momentum conveyed  by the torque in the disk; the  right hand side term is the  angular momentum transport  from the disk's surface by radiation. Using above two equations, one can find an expression for the energy flux F by eliminating  $ W_\phi^r$. Now, utilizing the relation between energy and angular momentum $dE = Q dJ$  for which the  circular geodesic orbits assumes the form $ E_{,r} = \Omega  L_{,r}  $,  one can  obtain the flux $ F$  of the radiant energy over the disk    in terms of the specific energy, angular momentum and of the angular velocity of the disk matter as,
\begin{equation}\label{eq:F}
F(r) =  - \frac{\dot{M}}{4 \pi  \sqrt{-g}}  \frac{\Omega_{,r}}{(E-\Omega L)^2}\int_{r_{ms}}^r (E-\Omega L) L_{,r} dr.
\end{equation}
Here,  the ``no torque" inner boundary condition was assumed  which  means torque disappears  at  the  inner edge of  the disk  as   the accreting matter at the marginally stable orbit $ r_{ms}$  drops spontaneously into the centre of the compact objects, and cannot wield significant torque on the disk. Note that since the radiant energy flux of the disk is  proportional to the accretion rate $\dot{M}$, hence the radiation flux from the disk increases linearly  with the accretion rate $\dot{M}$.

One more   significant feature of the mass accretion activity  is the efficiency  $\epsilon$  (  computed  by the specific energy of a particle at the marginally stable orbit $ r_{ms}$ when  all the radiated photons can seepage to infinity  )
with which the central object changes rest mass into outward radiation. The  accretion
efficiency $\epsilon$ (should be non-negative) is given by,
\begin{equation}
\epsilon = 1- E(r_{ms}).
\end{equation}
In the  steady-state thin disk accretion model, it is argued that  the  accreting  matter  is   in  thermodynamical  equilibrium.
As a result,  the radiation produced by the entire disk surface can be recounted as a perfect black body radiation, where the energy flux is given,
\begin{equation}
F (r) = \sigma  T^4(r),
\end{equation}
 with $ T$  is the effective   temperature       of the  geometrically thin  black-body disk, and $\sigma = 5.670374419 \times  10^{-8}~J~ s^{-1} m^{-2} K^{-4}$ is the Stefan-Boltzmann constant.
Allowing for that  the  disk radiates  as a  black body, then   the observed luminosity  $L (\nu)$  of the disk can be exemplified by a redshifted black body spectrum  as,
\begin{equation}
 L(\nu)  = 4 \pi d^2 I(\nu) = \frac{8 \pi h \cos i}{c^2} \int_{r_{in}}^{r_{out}}   \int_0^{2\pi} \frac{\nu_e^3 r  dr d\phi}{\exp[\frac{h \nu_e}{\kappa_B T}]-1}.
\end{equation}
Here, d is the distance  between the observer and the center of the disk ,  $ r_{in }$  and $ r_{out }$   stand for the inner and outer radii of the disc respectively  (  usually, one takes  $r_{in } = r_{ms }$ and $ r_{out } = \infty$       ) ,  $i $  is the disk inclination angle to the vertical,  $h$  is the Planck constant,     $k_B $   is  the  Boltzmann constant, $\nu_e $  is the emission frequency and  $  I(\nu) $ indicates  the thermal energy flux radiated by the disk. The observed photons are redshifted and the
detected  frequency  $\nu$  is associated with the discharged radiation
frequency  from the disk    as,
\begin{equation}
\nu_e = (1 + z) \nu.
\end{equation}

For a common axisymmetric metric  that represents  the space-time geometry around a rotating compact object,
 the redshift factor $(1 + z)$  which  comprises the effects
of both gravitational redshift and Doppler shift,   takes the following form,
\begin{equation}
1+z = \frac{1+\Omega r \sin i \sin \phi}{\sqrt{-g_{tt}- 2 \Omega g_{t \phi}- \Omega^2 g_{\phi \phi}}} ~~.
\end{equation}

Here,  the light bending effect has been neglected.

\section{Marginally Stable Circular Orbits :}

 The thin accretion disk  is made by the timelike particles moving  in stable circular timelike geodesics. These  timelike  geodesics around compact astrophysical objects are  very significant for theoretical as well as experimental point of view as they convey the information about the compact object where very strong gravity takes place.
For an arbitrary  stationary and axially symmetric geometry, let us consider
the metric   in a general form  as
\begin{equation}\label{eq:metric}
d s^2 =- g_{tt} d t^2 + 2  g_{t \phi} d t d \phi +  g_{r r} d r^2 +  g_{\theta \theta} d \theta^2+  g_{\phi \phi} d \phi^2.
\end{equation}

In  the equatorial approximation $ (|\theta-\frac{\pi}{2}|<<1)$,    the metric functions $g_{tt},  ~g_{rr}, g_{\phi \phi }, g_{t \phi} ~and~ g_{\theta \theta} ~$     will be the function of  the radial coordinate r.

Now we consider the geodesic of radial timelike particles  in the planes   $\theta = constant$,  for which    the Lagrangian assumes the  following form ( with $\tau$ is an affine parameter  of the world line of
the particle ),
\begin{equation}
L^*= - g_{tt} \dot{t}^2 + 2  g_{t \phi} \dot t \dot \phi +  g_{r r} \dot{ r}^2  +  g_{\phi \phi} \dot{ \phi}^2.
\end{equation}

Here the dot means derivative with respect to $\tau$.

Euler -Lagrangian equation  $ \left(
 \frac{d}{d\tau}\left(\frac{\partial L^*}{\partial \dot x}\right)- \frac{\partial L^*}{\partial x} = 0,~ ~x = t,   \phi \right) $
  yields,
\begin{align}
- g_{tt} \dot{t} +  g_{t \phi} \dot \phi =& -E,\nonumber\\
g_{t \phi} \dot t+  g_{\phi \phi} \dot{ \phi} = &L.
\end{align}

 Here, $E$  and $ L$  be  two constants of motion for particles  which are known as  the specific energy   and the specific angular momentum , respectively of the particles moving on circular orbits around the compact objects.
 Solving the above two equations , we obtain the orbiting particle's  four-velocity components as,
 \begin{align}
 u^t \equiv  \dot t&= \frac{L g _{t \phi}+ g_{\phi \phi}E}{g_{t t}g_{\phi \phi}+ g_{t \phi}^2},\nonumber\\
 u^\phi \equiv  \dot{\phi}&= \frac{- g _{t \phi} E + L g_{t t}}{g_{t t}g_{\phi \phi}+ g_{t \phi}^2}.
 \end{align}

For time-like particles  the   first integral of geodesics equation  ($g_{\mu \nu} \dot{x^\mu} \dot{x^\nu}  =  -1$)  yields,
\begin{eqnarray*}
  g_{r r} \dot{ r}^2 = -1 + g_{tt} \dot{t}^2 - 2  g_{t \phi} \dot t \dot \phi -  g_{\phi \phi} \dot{ \phi}^2.
\end{eqnarray*}
Putting the values of $\dot t$, $\dot{\phi}$, one obtains,
\begin{eqnarray*}
  g_{r r} \bigg (\frac{d r}{d \tau}\bigg)^2 = -1 +\frac{E^2 g_{\phi \phi}+ 2 E L g_{t \phi}- L^2 g_{t t}}{ g_{t \phi}^2+g_{t t}g_{\phi \phi}}.
\end{eqnarray*}
From the above equation, one can define an effective potential term as,
\begin{equation}\label{eq:V_eff}
V_{eff}(r)=-1 +\frac{E^2 g_{\phi \phi}+ 2 E L g_{t \phi}- L^2 g_{t t}}{ g_{t \phi}^2+g_{t t}g_{\phi \phi}}.
\end{equation}

For  stable  circular  orbits  in  the  equatorial  plane, the time-like geodesic satisfies $   \dot{r} =0,     \ddot{r} =0 ~and ~    \dddot{r}  < 0$. In    terms of the effective potential, these  conditions  take the forms  as ~$ V_{eff}  = 0, ~~ \frac{ d V_{eff}}{dr}   = 0~~ and ~~ \frac{ d^2 V_{eff}}{dr^2}    < 0.  $
The first two conditions     provide the kinematic parameters such as the  specific energy,  the specific angular momentum and the angular velocity  of the particles moving in the stable circular orbits .

Here,
\begin{eqnarray*}
\frac{d \phi}{d t}=    \frac{\dot{\phi}}{\dot{t}} = \Omega = \frac{ L g_{t t}- g _{t \phi} E }{L g _{t \phi}+ g_{\phi \phi}E}.
\end{eqnarray*}
\begin{equation}\label{eq:L}
\Rightarrow L= \frac{  (g_{t \phi}+\Omega g _{\phi \phi}) E }{ g _{t t}- \Omega  g_{t \phi}}.
\end{equation}
Now, $V_{eff}(r)=0  $ implies
\begin{eqnarray*}
 E^2 g_{\phi \phi}+2  E L  g_{t \phi} -L^2 g_{tt} =( g_{t \phi})^2 +  g_{t t}  g_{\phi \phi} .
\end{eqnarray*}
Putting the value of $L$ in the above equation, from eq. (\ref{eq:L}) one can get
\begin{equation}\label{eq:E}
E = \frac{g _{t t}- \Omega  g_{t \phi}}{\sqrt{g_{tt}- 2 \Omega g_{t \phi}- g_{\phi \phi } \Omega^2}}.
\end{equation}
So, using the  above expression in eq. (\ref{eq:L}) we obtain,
\begin{equation}\label{eq:LL}
L= \frac{g _{t \phi}+\Omega  g_{\phi \phi}}{\sqrt{g_{tt}- 2 \Omega g_{t \phi}- g_{\phi \phi } \Omega^2}}.
\end{equation}

 The  condition, $\frac{d V_{eff}(r)}{dr} =0$,  yields,
\begin{equation}\label{eq:Omega}
 \Omega = \frac{-g_{t \phi, r} \pm \sqrt{(g_{t \phi,r})^2+ g_{t t,r}  g_{\phi \phi,r}}}{g_{\phi \phi,r}}.
\end{equation}
The marginally stable circular orbits around the  compact object  could be found    from the following condition:
\begin{equation}
\frac{ d^2 V_{eff}}{dr^2}    = 0.
\end{equation}

\section{Accretion disk properties of a rotating traversable wormhole of Teo class}
We first study a rotating generalization of the static Morris-Thorne wormhole \cite{kg14}.
For that consider the following stationary, axisymmetric spacetime metric describing a rotating traversable wormhole of Teo class \cite{Teo}, given as,
\begin{equation}\label{eq:metric1}
ds^2=-N^2dt^2+\frac{dr^2}{1-\frac{b}{r}}+r^2K^2\left[ d\theta^2+\sin^2\theta\left(d\varphi-\omega dt^2\right) \right],
\end{equation}
 where $-\infty <t < \infty$, and $r_0 \leq r < \infty$ ,$ 0 \leq \theta \leq \pi$ and $0 \leq \varphi \leq 2\pi$ are in spherical coordinates.
$N, b, K$, and $\omega$ are functions of $r$ and $\theta$ only, such that they are regular on the axis of  symmetry $\theta = 0, \pi$. This spacetime describes two identical, asymptotically flat regions, with the throat, $ r = r_0 = b > 0$ as the boundary. Since the spacetime geometry of the two regions of the wormhole connected through the throat are exact copies, the wormhole is symmetric about the throat.
 The function $N$ is interpreted as the redshift function while $b$ as the shape function of the wormhole. In order to represent a wormhole the metric eq.(\ref{eq:metric1}) must be devoid of any event horizons or curvature singularities.
For that the redshift function N must be non-zero and finite everywhere in $r_0\leq r<\infty$.  Also we must have, $\partial_{\theta} b\mid_{r = r_0}=0$ to avoid any event horizon or curvature singularity at the throat, and  the shape function $b$ must satisfy the flare-out condition at the throat, $\partial_r b\mid_{r=r_0}<1$ and $b\leq r$.
The function $K$ determines the area radius given by $R = rK$ while the function $\omega$ measures the angular velocity of the wormhole \cite{Teo}. As an specific example, we consider a wormhole considered in \cite{Konoplya_2016} with the metric functions given by: \begin{align}
N&=\sqrt{1-\frac{2M}{r}},\nonumber\\
b&=\frac{r_0^2+2M\left(r-r_0\right)}{r},\nonumber\\
K&=1,\nonumber\\
\omega&=\frac{2M a}{r^3}=\frac{2J}{r^3}.
\end{align}

where J is the angular momentum of the wormhole and $M$ is the mass. The throat is at $r=r_0>2 M$. \par

From eq.(\ref{eq:V_eff}) we can write the expression of effective potential as :
\begin{equation}
V_{eff}=\frac{-E^2r^6+4ELJr^3-2L^2Mr^3+L^2r^4-2r^5M+r^6-4J^2L^2}{r^5\left(-r+2M \right)}  ,
\end{equation}
which is obtained from the geodesic motion of the test particles in the
equatorial plane of the wormhole.
From the conditions, $V_{eff}(r)=0$ and $\frac{dV_{eff}(r)}{dr}=0$, the expressions of the various kinematic properties can be obtained as below. From eq. (\ref{eq:E}) we get the specific energy as,
\begin{equation}
E=\frac{r^4-2Mr^3-6J^2+2J\sqrt{Mr^3+9J^2}}{r^2\sqrt{r^4-3Mr^3-18J^2+6J\sqrt{Mr^3+9J^2}}},
\end{equation}
and from eq. (\ref{eq:LL}) we obtain the specific angular momentum as,
\begin{equation}
L=\frac{r\left(\sqrt{Mr^3+9J^2}-3J\right)}{\sqrt{r^4-3Mr^3-18J^2+6J\sqrt{Mr^3+9J^2}}}.
\end{equation}
The angular velocity of the particles in the accretion disk, moving in the stable circular orbit are given by eq. (\ref{eq:Omega}) as,
\begin{equation}
\Omega=\frac{\sqrt{Mr^3+9J^2}-J}{r^3}.
\end{equation}

\begin{figure}[htbp]
\centering
\subfigure[]{\includegraphics[scale=0.65]{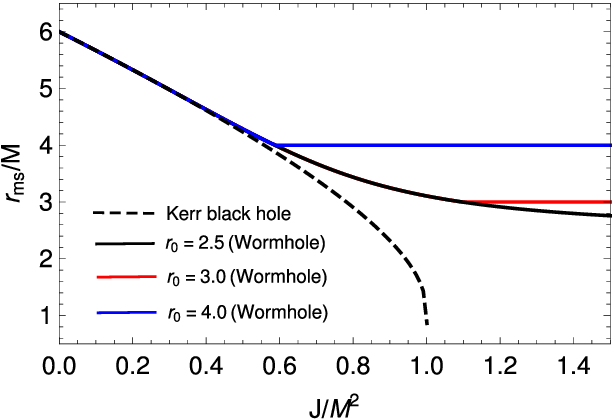}}
\subfigure[]{\includegraphics[scale=0.7]{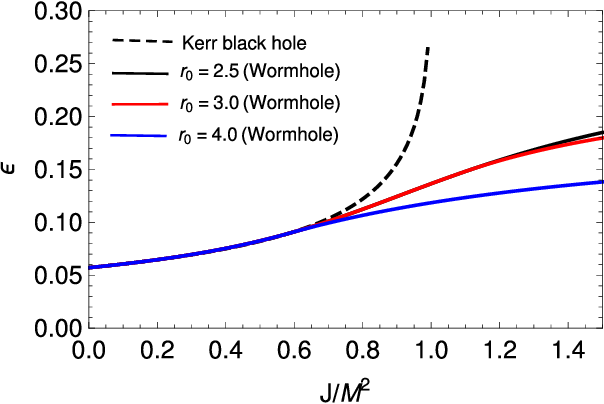}}
\caption{The Marginally stable circular orbits ($r_{ms}$) and the efficiency of an accretion disk ($\epsilon$), for different values of $J/M^2$ and $r_0$.}
\label{fig:rms}
\end{figure}

In Fig. \ref{fig:rms}, we have shown the dependence of the marginally stable circular orbits $r_{ms}$ and the efficiency $\epsilon$ of the accretion disk on of the angular momentum of the wormhole $J$ for different wormhole throat size $r_0$, along with the corresponding results for a Kerr black hole. From a physical point of view, the marginally stable circular orbits form the innermost stable circular orbits for the wormhole, beyond which the particle will fall into the throat and will be lost to the other universe. Note that, for a given angular momentum and throat size, $r_{ms}$ cannot be less than $r_0$. If all the roots of $\frac{d^2V_{eff}}{dr^2}=0$ are found to be smaller than $r_0$, then we take $r_{ms}=r_0$ as the throat acts as the position of the marginal stable orbits in that case. See \citep{Paul} for more details.  In Table \ref{tab:rms}, the numerical values of the position of marginally stable circular orbits $r_{ms}$ and the efficiency $\epsilon$ of the accretion disk, corresponding to various values of the angular momentum of the wormhole $J$ and throat size $r_0$, are tabulated, along with the corresponding results for a Kerr black hole.

\begin{table}[h!]
\centering
\caption{The Marginally stable circular orbits ($r_{ms}$) and the efficiency of the accretion disk ($\epsilon$), for different values of $J/M^2$ and $r_0$.}
 \begin{tabular}{| c | c | c | c | c | c | c|}
 \hline\hline
  & \multicolumn{3}{c|}{$r_{ms}$ (in M)} &  \multicolumn{3}{c|}{Efficiency $\epsilon$} \\
  & \multicolumn{2}{c|}{Wormhole} & Kerr Black &  \multicolumn{2}{c|}{Wormhole} & Kerr Black  \\
  $J/M^2$ & $r_0=2.5M$ & $r_0=4M$ & hole & $r_0=2.5M$ & $r_0=4M$ & hole\\
 \hline
  0 & 6.0 & 6.0 & 6.0 & 0.05719 & 0.05719 & 0.05719 \\
  0.2 & 5.32761 & 5.32761 & 5.32944 & 0.06474 & 0.06474 & 0.064634 \\
  0.5 & 4.28536 & 4.28536 & 4.23300 & 0.08254 & 0.08254 & 0.08212 \\
  0.7 & 3.67576 & 4.0 & 3.39313 & 0.10112 & 0.09940 & 0.10361 \\
  0.9 & 3.25027 & 4.0 & 2.32088 & 0.12446 & 0.11299 & 0.15575 \\
  1.5 & 2.75805 & 4.0 & --- & 0.18496 & 0.13821 & ---\\
 \hline\hline
 \end{tabular}
\label{tab:rms}
\end{table}

\begin{figure}[htbp]
\centering
\includegraphics[scale=0.63]{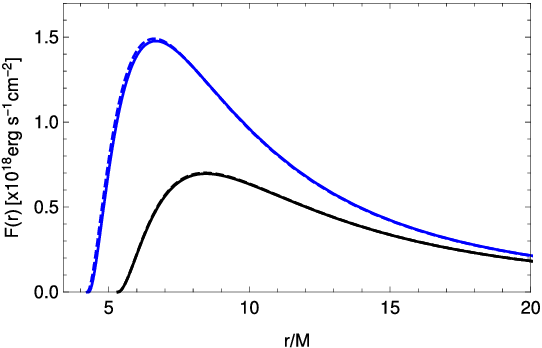}
\includegraphics[scale=0.63]{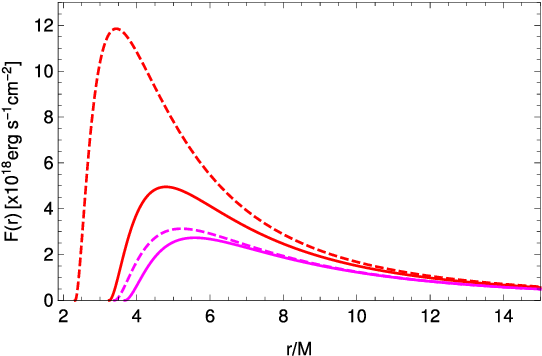}
\includegraphics[scale=0.63]{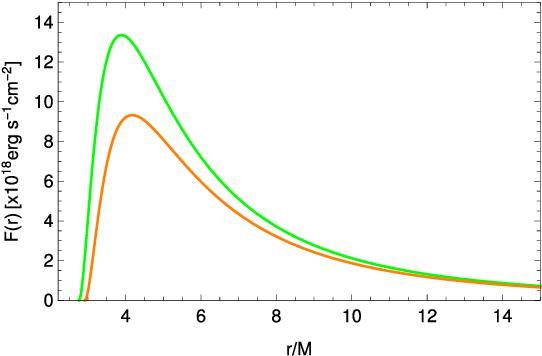}
\includegraphics[scale=0.63]{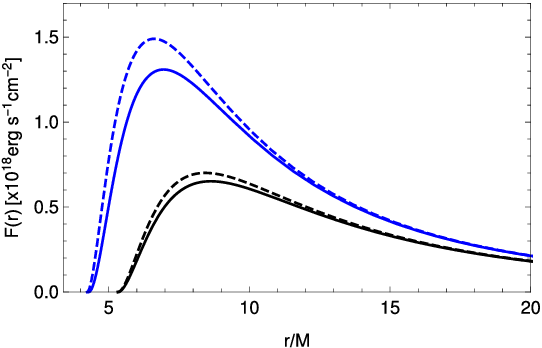}
\includegraphics[scale=0.63]{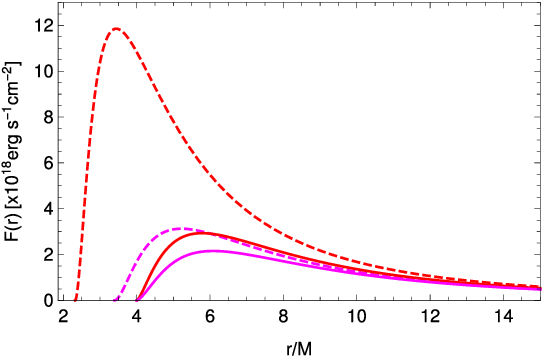}
\includegraphics[scale=0.63]{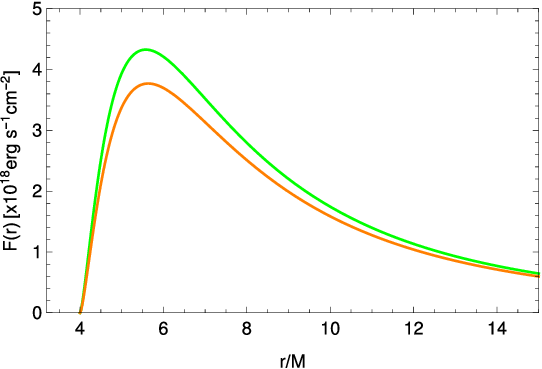}
\caption{The energy flux radiated by the accretion disk around the wormhole (solid) for dimensionless spin values $J/M^2=0.2$ (black), $0.5$ (blue), $0.7$ (magenta), $0.9$ (red), $1.2$ (orange) and $1.5$ (green). The first and second row, respectively, corresponds to the throat size $r_0=2.5M$ and $4M$. For comparison, we also have shown the corresponding plots for a Kerr black hole (dashed) of same mass and spin. Some solid and dashed plots coincide with one another.}
\label{fig:F}
\end{figure}

\begin{figure}[htbp]
\centering
\includegraphics[scale=0.63]{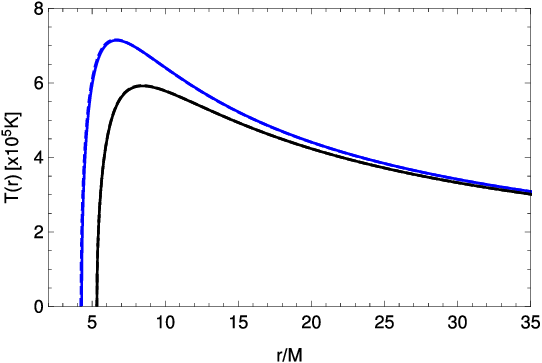}
\includegraphics[scale=0.63]{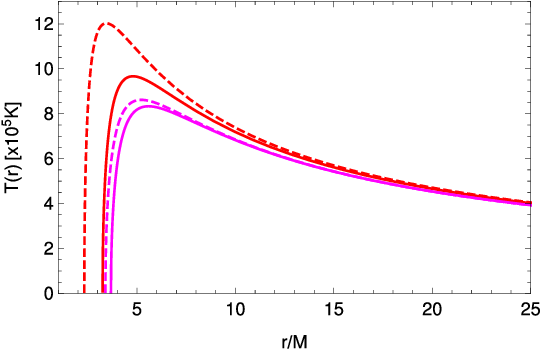}
\includegraphics[scale=0.63]{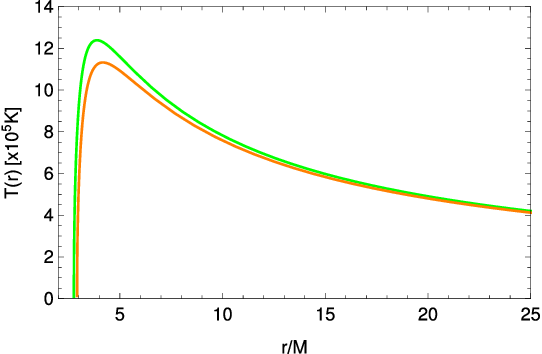}
\includegraphics[scale=0.63]{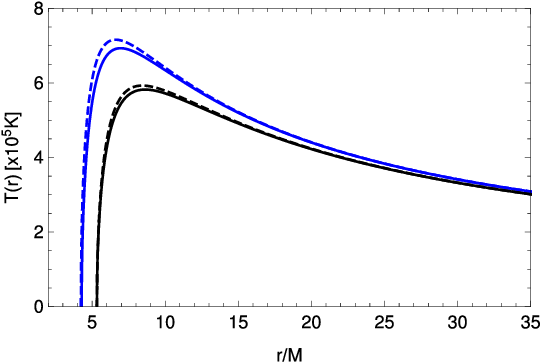}
\includegraphics[scale=0.63]{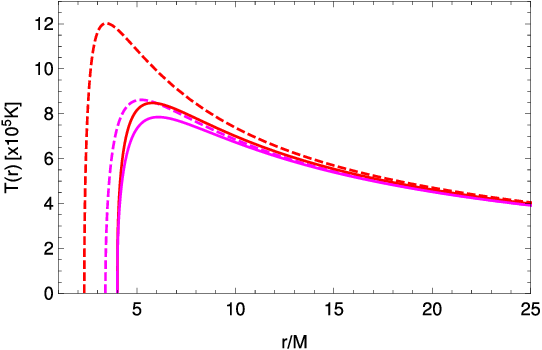}
\includegraphics[scale=0.63]{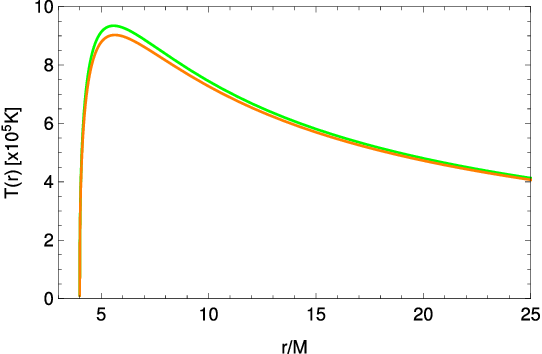}
\caption{The temperature distribution of the accretion disk around the wormhole (solid) for dimensionless spin values $J/M^2=0.2$ (black), $0.5$ (blue), $0.7$ (magenta), $0.9$ (red), $1.2$ (orange) and $1.5$ (green). The first and second row, respectively, corresponds to the throat size $r_0=2.5M$ and $4M$. For comparison, we also have shown the corresponding plots for a Kerr black hole (dashed) of the same mass and spin. Some solid and dashed plots coincide with one another.}
\label{fig:T}
\end{figure}

\begin{figure}[htbp]
\centering
\includegraphics[scale=0.63]{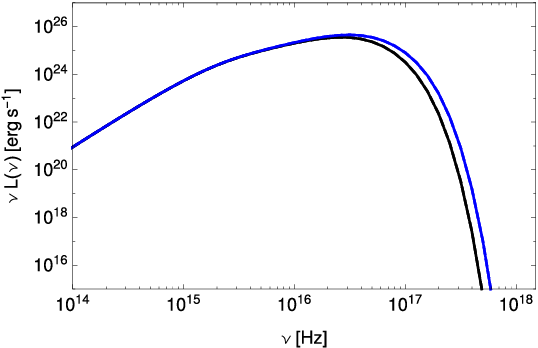}
\includegraphics[scale=0.63]{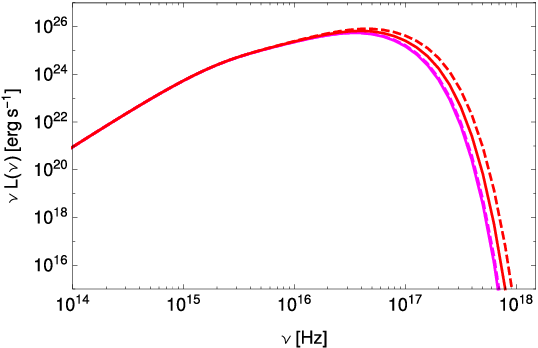}
\includegraphics[scale=0.63]{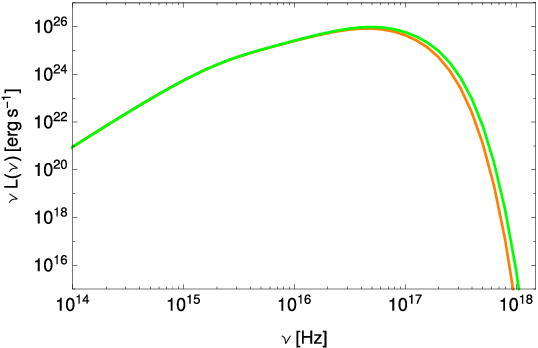}
\includegraphics[scale=0.63]{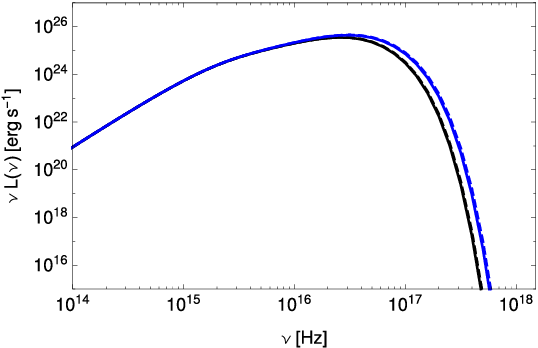}
\includegraphics[scale=0.63]{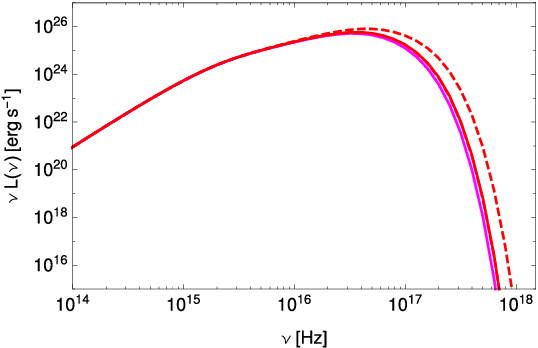}
\includegraphics[scale=0.63]{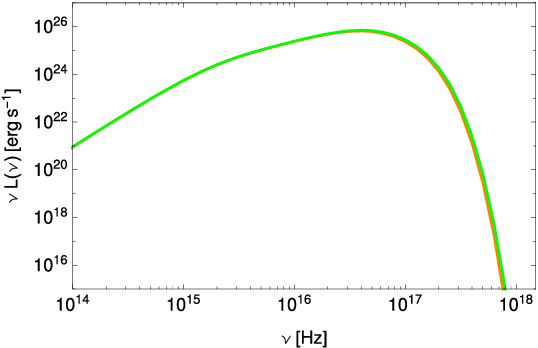}
\caption{The emission spectra of the accretion disk around the wormhole (solid) for dimensionless spin values $J/M^2=0.2$ (black), $0.5$ (blue), $0.7$ (magenta), $0.9$ (red), $1.2$ (orange) and $1.5$ (green). The first and second row, respectively, corresponds to the throat size $r_0=2.5M$ and $4M$. For comparison, we also have shown the corresponding plots for a Kerr black hole (dashed) of the same mass and spin. Some solid and dashed plots coincide with one another.}
\label{fig:lum}
\end{figure}

To obtain the  flux radiated, temperature and emission spectra of the disk, we consider the mass of the wormhole to be $M=15 M_{\odot}$ \citep{Imp} and the mass accretion rate to be $\dot{M} \approx 3\times 10^{-12}M$/year. Also, to obtain the results in the physical units, we restore $G$ and $c$ by replacing $M$ by $GM/c^2$ and $\dot{M}$ by $\dot{M}c^2$, where $G$ and $c$ are, respectively, Newton's gravitational constant and speed of light in vacuum. The flux $F(r)$ and the temperature $T(r)$ of the disk for different values of the dimensionless spin $J/M^2$ and throat size $r_0$ are, respectively, plotted in Figs. \ref{fig:F} and \ref{fig:T}, along with the corresponding results for a Kerr black hole. From the figures, we can note that the flux and the temperature sharply increase just near the throat $r_0$ and then decrease gradually. Similarly, the emission spectra of the disk for various values of the dimensionless spin $J/M^2$ is plotted in Fig. \ref{fig:lum}, along with the emission spectra of a Kerr black hole. Note that, when both the spin and throat size are small, the flux, the temperature and the emission spectra of the disk mimic those of a Kerr black hole. However, with increasing either the spin value or the throat size, the results for the wormhole start differing from those of a Kerr black hole. In Table \ref{tab:nu}, following \citep{Imp}, we have shown the critical frequency $\nu_{crit}$ at which the spectra of the disk have a maximum values $\nu L(\nu)$.

\begin{table}[h!]
\centering
\caption{The critical frequency $\nu_{crit}$ at which the spectra has a maximum value $\nu_{crit}L_{max}(\nu_{crit})$ for different values of $J/M^2$ and $r_0$.}
 \begin{tabular}{| c | c | c | c | c | c | c|}
 \hline\hline
  & \multicolumn{3}{c|}{$\nu_{crit}$($\times 10^{16}$ Hz)} &  \multicolumn{3}{c|}{$\nu_{crit}L_{max}(\nu_{crit})$ ($\times 10^{25} $erg $\text{s}^{-1}$)} \\
  & \multicolumn{2}{c|}{Wormhole} & Kerr Black &  \multicolumn{2}{c|}{Wormhole} & Kerr Black  \\
  $J/M^2$ & $r_0=2.5M$ & $r_0=4M$ & hole & $r_0=2.5M$ & $r_0=4M$ & hole\\
 \hline
  0.2 & 2.663 & 2.631 & 2.665 & 3.608 & 3.559 & 3.612 \\
  0.5 & 3.110 & 3.048 & 3.110 & 4.544 & 4.433 & 4.539 \\
  0.7 & 3.520 & 3.387 & 3.588 & 5.499 & 5.216 & 5.648 \\
  0.9 & 3.975 & 3.644 & 4.503 & 6.683 & 5.869 & 8.195 \\
  1.2 & 4.555 & 3.904 & --- & 8.432 & 6.618 & --- \\
  1.5 & 4.942 & 4.072 & --- & 9.769 & 7.168 & --- \\
 \hline\hline
 \end{tabular}
\label{tab:nu}
\end{table}

\section{Images of an accretion disk around the wormhole}

\begin{figure}[ht]
\centering
\subfigure[~$J/M^2=0.2$ (Kerr black hole)]{\includegraphics[scale=0.55]{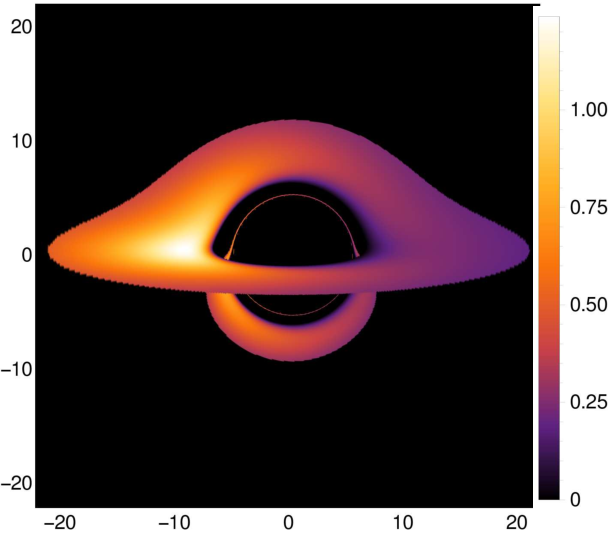}\label{fig:disk_image_SS_a}}
\subfigure[~$J/M^2=0.5$ (Kerr black hole)]{\includegraphics[scale=0.55]{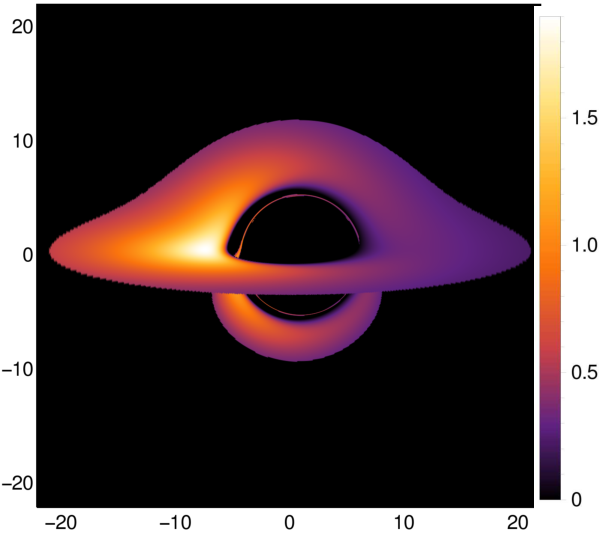}\label{fig:disk_image_SS_b}}
\subfigure[~$J/M^2=0.9$ (Kerr black hole)]{\includegraphics[scale=0.55]{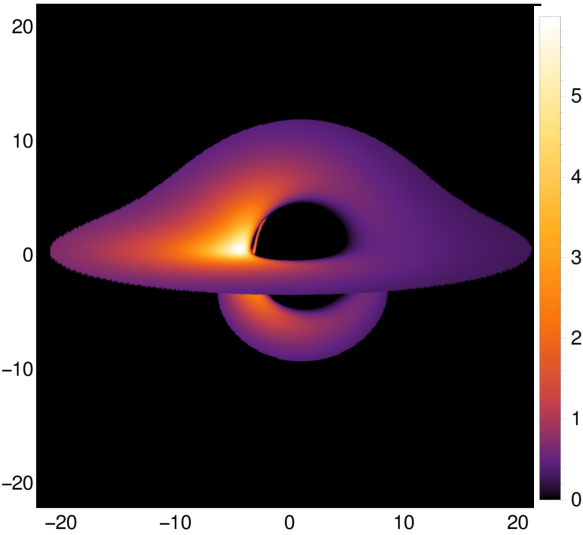}\label{fig:disk_image_SS_c}}
\subfigure[~$J/M^2=0.2$, $r_0=2.5$ (Wormhole)]{\includegraphics[scale=0.55]{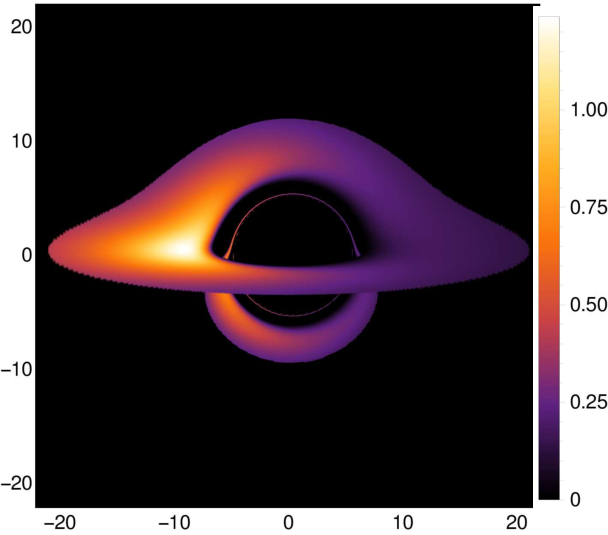}\label{fig:disk_image_SS_d}}
\subfigure[~$J/M^2=0.5$, $r_0=2.5$ (Wormhole)]{\includegraphics[scale=0.55]{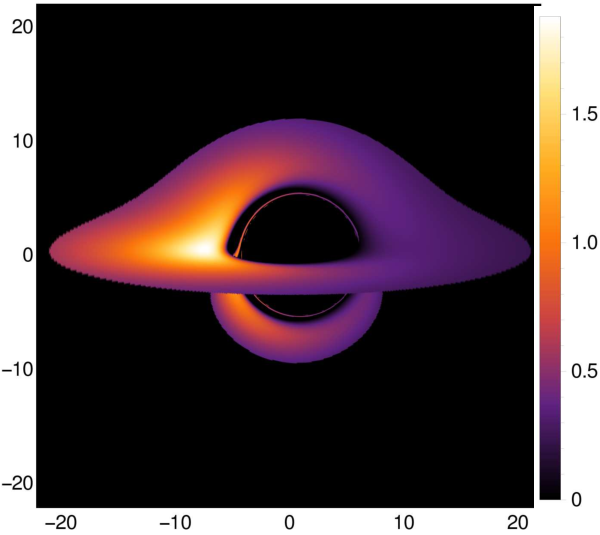}\label{fig:disk_image_SS_e}}
\subfigure[~$J/M^2=0.9$, $r_0=2.5$ (Wormhole)]{\includegraphics[scale=0.55]{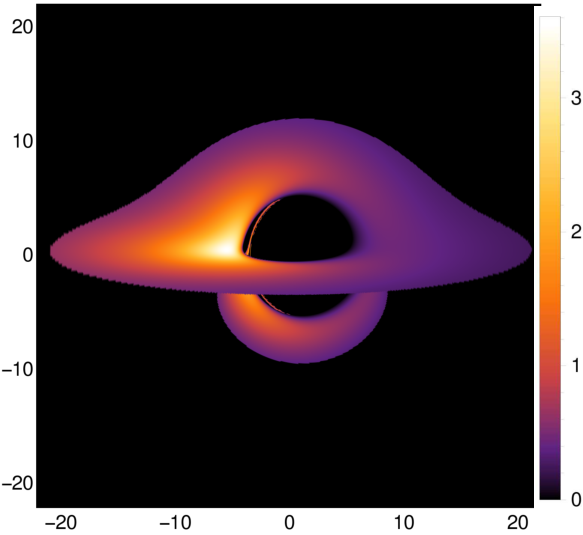}\label{fig:disk_image_SS_f}}
\subfigure[~$J/M^2=0.5$, $r_0=3.0$ (Wormhole)]{\includegraphics[scale=0.55]{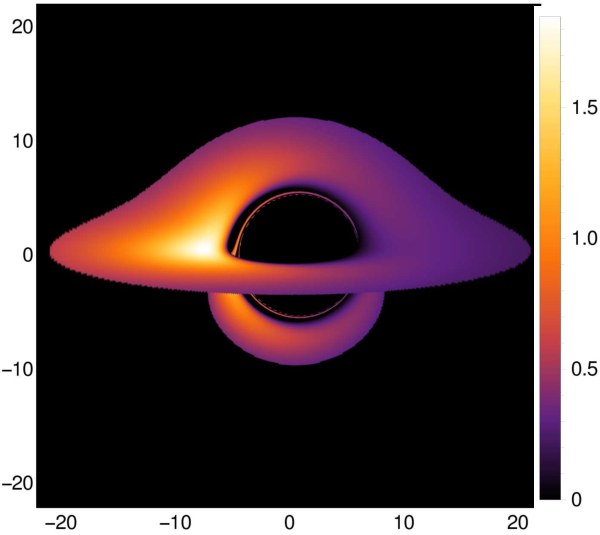}\label{fig:disk_image_SS_g}}
\subfigure[~$J/M^2=0.5$, $r_0=4.0$ (Wormhole)]{\includegraphics[scale=0.55]{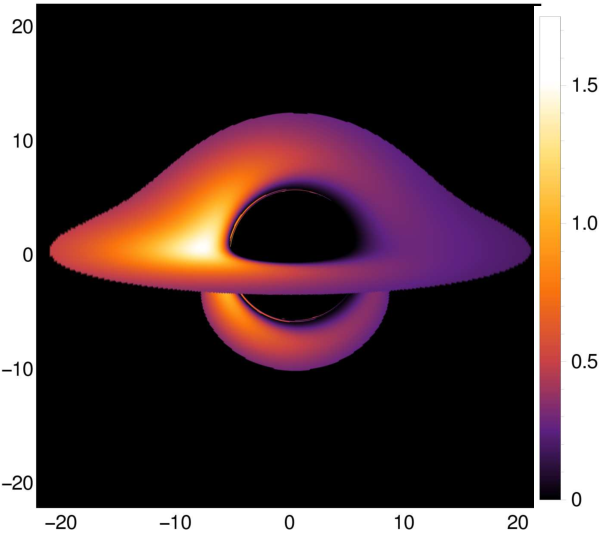}\label{fig:disk_image_SS_h}}
\subfigure[~$J/M^2=0.5$, $r_0=5.0$ (Wormhole)]{\includegraphics[scale=0.55]{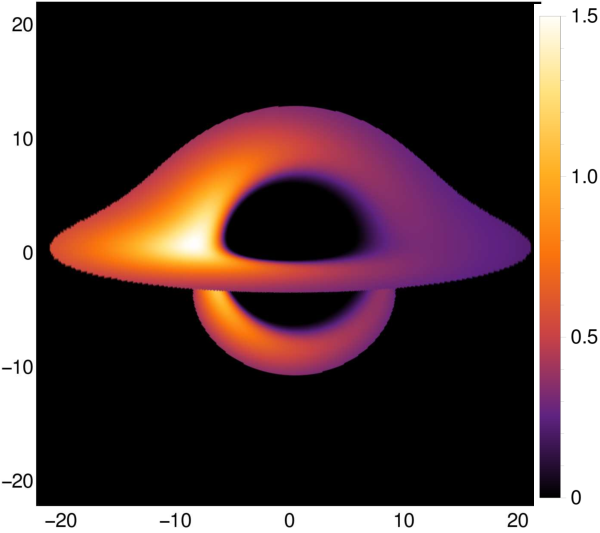}\label{fig:disk_image_SS_i}}
\caption{The images of an accretion disk in a Kerr black hole and rotating wormhole [(a)-(i)]. For wormhole, the disk is on the same side of the observer. The outer edge of the disk is at $r = 20M$ , and the observer's inclination angle is $\theta_{o}=80^{\circ}$. The observer is placed at the radial coordinate $r=10^4M$, which corresponds effectively to the asymptotic infinity. In order to get rid of the parameters $M$ and $\dot{M}$, we have normalized the fluxes by the maximum flux observed for the Schwarzschild black hole. Also, we have plotted square-root of the normalized flux for a  better visualization. The color bars show the values the flux. All spatial coordinates are in units of $M$.}
\label{fig:disk_image_SS}
\end{figure}

\begin{figure}[ht]
\centering
\subfigure[~$J/M^2=0.2$, $r_0=2.5$ (Wormhole)]{\includegraphics[scale=0.55]{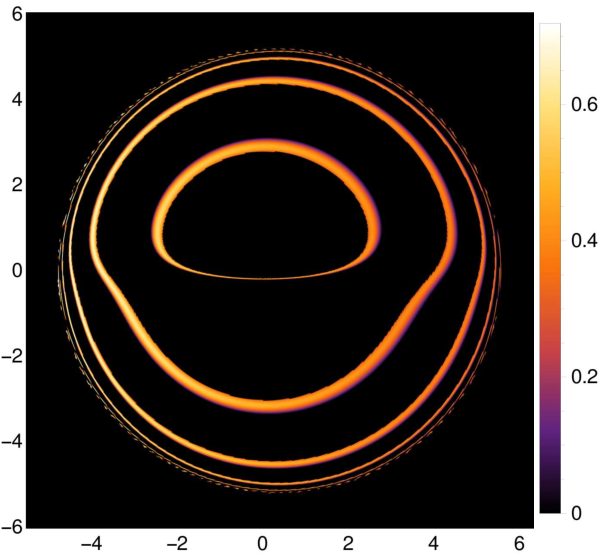}}
\subfigure[~$J/M^2=0.5$, $r_0=2.5$ (Wormhole)]{\includegraphics[scale=0.55]{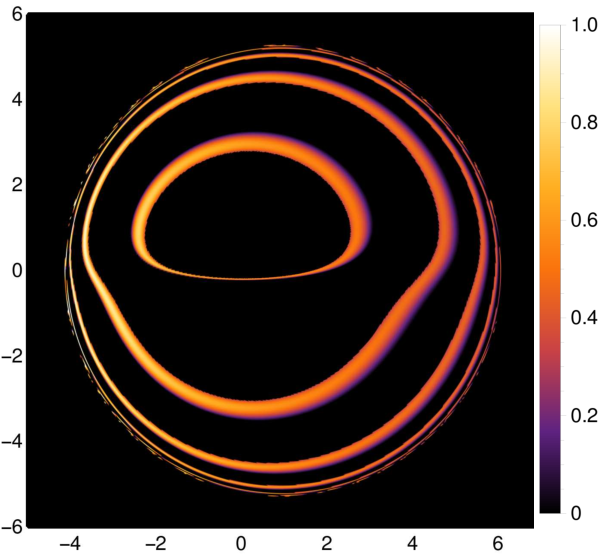}}
\subfigure[~$J/M^2=0.9$, $r_0=2.5$ (Wormhole)]{\includegraphics[scale=0.55]{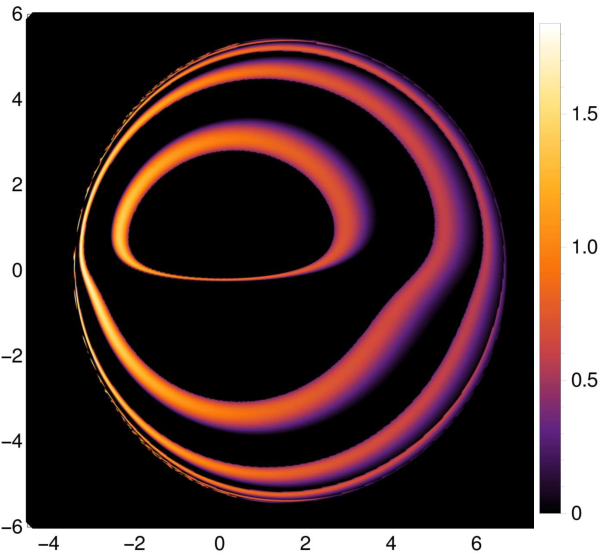}}
\caption{The images of an accretion disk in the rotating wormhole [(a)-(c)] when the disk is on the other side of the throat (i.e. the disk and the observer are on two opposite sides of the throat). The outer edge of the disk is at $r = 20M$ , and the observer's inclination angle is $\theta_{o}=80^{\circ}$. The observer is placed at the radial coordinate $r=10^4M$, which corresponds effectively to the asymptotic infinity. In order to get rid of the parameters $M$ and $\dot{M}$, we have normalized the fluxes by the maximum flux observed for the Schwarzschild black hole. Also, we have plotted square-root of the normalized flux for better looking. The color bars show the values of the flux. All spatial coordinates are in units of $M$.}
\label{fig:disk_image_OS}
\end{figure}

We now consider how the image of an accretion disk around the wormhole looks like. In order to produce the intensity map of the image, we implement ray-tracing techniques discussed in \citep{Paul,Shaikh_disk}. We do not discuss the detailed procedure here as it can be found in \citep{Paul,Shaikh_disk}. Figure \ref{fig:disk_image_SS} shows how the image of an accretion disk looks like to an asymptotic observer when the disk and the observer are on the same side of the wormhole throat. For comparison, we also have shown the images for a Kerr black hole. Note that, when   both the spin and the wormhole throat size are small, the images almost mimic those of the black hole (compare among Figs. \ref{fig:disk_image_SS_a},\ref{fig:disk_image_SS_b}, \ref{fig:disk_image_SS_d} and \ref{fig:disk_image_SS_e}). However, with increasing spin value or the throat size, the image starts differing both qualitatively and quantitatively from that of a Kerr black hole (compare the flux values and the inner thin rings in Figs. \ref{fig:disk_image_SS_c} and \ref{fig:disk_image_SS_f} and in Figs. \ref{fig:disk_image_SS_b}, \ref{fig:disk_image_SS_h} and \ref{fig:disk_image_SS_i}). This indicates that, when the disk and the observer are on the same side, wormholes with larger spin or throat size can be distinguished from a Kerr black hole with the same spin through the observation of their disk images. Figure \ref{fig:disk_image_OS} shows the image of an accretion disk when the disk and the observer are on the two opposite sides of the wormhole throat. Note that, irrespective of the spin or the throat size, the images in this case are strikingly different from those of a Kerr black hole. In this case, a single intensity map of a disk image contains multiple images. This unique characteristic feature of the images can in principle be used to distinguish a wormhole from a black hole.

\section{Concluding Remarks}
In this paper, we have investigated  thin accretion disk  around a rotating traversable wormhole of Teo class \cite{Teo}, of mass $M=15 M_{\odot}$. We have studied the variation of the physical properties of our wormhole, with the spin parameter $J/M^2$ ranging from 0.2 to 1.5 and for two specific values of throat radius: $r_0=2.5M $ and $r_0=4M$; and hence compared those results with that of Kerr black hole. In classical theory of general relativity, traversable wormholes are supported by exotic matter in their throat, which is characterized by a stress energy tensor that violates the null energy condition. This unique feature may help to distinguish it and compare it with the Kerr black hole geometry. The expressions of the basic kinematic properties of the accretion disk, like angular velocity $\Omega$, specific energy $E(r)$, specific angular momentum $L(r)$ and effective potential term $V_{eff}$, are  investigated analytically. The effect of changing the spin parameter $J/M^2$ on the marginally stable circular orbits $r_{ms}$ are studied graphically in figure 1.(a) corresponding to three values of throat radius: $r_0=2.5 M$ \textit{(black)},$3.0 M $ \textit{(red)} and $ 4.0 M$ \textit{(blue)}; compared to the dashed curve representing the Kerr black hole.  Note that the curves are almost overlapping for small values of $J/M^2$, but as $J/M^2$ increases beyond $0.5$ they separate out, though the value of $r_{ms}$ can never be less than the throat radius $r_0$, which explains the fact that each of the curve corresponding to different throat radius ends in a straight line with the constant value of $r_{ms}=r_0$ . The values of $r_{ms}$ are found to be decreasing with increase in $J/M^2$, but are more than the values obtained in case of Kerr black hole. The efficiency of accretion $\epsilon$ is plotted in figure 1(b). In general, for all the different values of throat radius $r_0$, efficiency increases with increase in $J/M^2$, but is found to be lesser than the Kerr black hole. For small value of $J/M^2$ the curves overlap but for $J/M^2> 0.5$, we find that the efficiency for the wormhole decreases for increasing values of throat radius $r_0$.  These features of marginally stable circular orbits $r_{ms}$ and efficiency $\epsilon$ can also be studied in details from table I, corresponding to the different $J/M^2$ values and for the two specific throat radius $r_0=2.5 M$ and $4.0 M$ . Note that the efficiency $\epsilon$ of  accretion  ranges between 5.7\% to 18.5 \% approximately.\par
Now to analyze the properties of the radiation emerging from the surface of the disk, we have plotted the energy flux of perfect black body radiation $F(r)$ [in $erg$ $sec^{-1} cm^{-2}$] and temperature distribution $T(r)$ [in $K$], in figures 2 and 3, respectively, for six different values of the dimensionless spin parameter $J/M^2$ ranging from 0.2 to 1.5 and for throat $r_0=2.5 M$ \textit{(along first row)} and $r_0=4 M$ \textit{(along second row)} and compared those with that of Kerr black hole \textit{(in dashed curves)}.  The radiation flux is of the order of $10^{18}$ $erg sec^{-1} cm^{-2}$ while temperature is of order $10^5 K$. In figure 4, the emission spectra $\nu L(\nu)$, [in $erg s^{-1}$] of the accretion disk around the wormhole is plotted corresponding to the different values of $J/M^2$ and $r_0$ as mentioned before.  {In table II, we have tabulated the value of frequency, called the  critical frequency $\nu_{crit}$ where the emission spectra has a maximum value of luminosity of the disk $\nu_{crit} L_{max}(\nu_{crit})$, corresponding to the different values of the dimensionless spin parameter $J/M^2$ and the throat radius $r_0$; in order to understand in details the effect of these two parameters on the luminosity of the disk. From the table it is evident that, along each column corresponding to the two throat radius $r_0$ values, the critical frequency $\nu_{crit}$ increases with increase in spin parameter $J/M^2$; but along any row, corresponding to any particular value of $J/M^2$, the $\nu_{crit}$ value is slightly higher in case of lower value of throat radius $r_0$. Same trend is noted in case of the maximum luminosity values $\nu_{crit} L_{max}(\nu_{crit})$ column. It increases with the increase in $J/M^2$ along each column, when the $r_0$ values are kept constant; but along any particular row, corresponding to a constant value of $J/M^2$, the values are higher in case of the lower value of $r_0$. Also we can find that for lower values of both spin parameter $J/M^2$ and throat $r_0$, the critical frequency and maximum luminosity $\nu_{crit} L_{max}(\nu_{crit})$ are closer to the Kerr black hole case, but their difference gradually increases when the two parametric values increases.} \par
In section V, we have tried to visualize the image of an accretion disk around the wormhole, by the intensity map of the image using ray-tracing techniques discussed in \citep{Paul,Shaikh_disk}. Figures 5.(d)-(i) shows how the image of an accretion disk looks like to an asymptotic observer when the disk and the observer are on the same side of the wormhole throat. Note that for smaller values of both the spin and the wormhole throat size , the images almost mimic those of the Kerr black hole shown in figures 5. (a)-(c), but when the spin parameter values or the throat radius is increased, the image starts differing qualitatively and quantitatively from that Kerr black hole. In figures 6.(a)-(c), the  images of an accretion disk in the rotating wormhole are shown, as observed by an asymptotic observer , when the disk is on the other side of the throat (i.e.the disk and the observer are on two opposite sides of the throat). Note that, irrespective of the spin or the throat size, the images of the accretion disk when the observer is located at the opposite side of the throat, are strikingly different from those of a Kerr black hole; and a single intensity map of a disk image contains multiple images. This unique characteristic feature of the images can in principle be used to distinguish a wormhole from a black hole. Thus the detailed analysis of the physical and geometrical properties of these images, studied in this paper, could yield invaluable information on the nature of the accretion flows, in case of a rotating wormhole. Also the striking results from image analysis of its emission spectra could help to detect and distinguish wormhole geometries from other compact objects. We hope our work could be an interesting study and could motivate similar research in this field.   {For the Kerr black hole geometry, the horizons exist for $a<M$ case. For $a>M$, the horizons do not exist, and we have a Kerr naked singularity. While we hope that future observations might tell whether or not a Kerr naked singularity exists, on the theoretical front, there has been a lot of works on what could be observational aspects of a Kerr naked singularity if it exists in nature
\citep{new1,new2,new3,new4,new5}. Moreover, in presence of other parameters (or hairs), a spacetime can represent a black hole even when $a>M$. For example, a tidally charged spacetime with $a>M$ can represent a black hole  [ see   \citep{new6} for a work on its shadow in light of M87$^*$ ]. Therefore, some black holes with parameters other than mass and spin or some horizonless object might have  $a>M$, although their existence and otherwise should be confirmed through future observations.}

\pagebreak

\section*{Acknowledgments}

FR would like to thank the authorities of the Inter-University Centre for Astronomy and Astrophysics, Pune, India for providing research facilities. This work  is a part of the project submitted by FR in SERB, DST. We are also thankful to the referees for their valuable comments and constructive suggestions.

\section*{Appendix}

One can note that our Lorentzian  wormhole has   a throat at $r = r_0$. When $M = 0$,  we have the massless wormhole  found by Ellis and Bronnikov \citep{new7,new8}. Also one can check that   , $\partial_\theta b |_{r=r_0} = 0$. Thus no  event horizon or curvature singularity at the throat. For this choice of b the flare out condition $\partial_rb |_{r=r_0} < 1$ and $ b < r$ have  been satisfied. The wormhole considered in eq. 27 was considered in reference [27] of our paper to study quasinormal ringing of a
wormhole. It has been shown that, depending on the parameters, the wormhole may ring similarly or completely
differently from that of a Kerr black hole. We took this metric to explore it in the context of accretion disk properties.

To check the traversability of this wormhole, let us find out the tidal acceleration at the throat. It shall be less than one earth gravity, i.e $ \leq 9.8 m/s^2$
 for traversable. But, for  simplicity, we consider the spin zero case below.
\begin{eqnarray*}
a^{\hat{j'}} &=& - c^2 R^{\hat{j}'}_{\hat{0}'\hat{k}'\hat{0'}}\xi^{\hat{k}}  \textrm{[ see eq. 46 Morris and Thorne paper in reference [2]   ]}\\'
a'&=& - c^2 R^{\hat{1}'}_{\hat{0}'\hat{1}'\hat{0}'}\xi^{\hat{1}} \textrm{( radial acceleration)}.
\end{eqnarray*}
Now,
\begin{eqnarray*}
R^{\hat{1}'}_{\hat{0}'\hat{1}'\hat{0}'} = -\left(1-\frac{b}{r}\right)\bigg[ - \phi'' + \frac{ b' r - b}{ 2 r (r - b)} \phi' -{\phi}'^2 \bigg].
\end{eqnarray*}
Now,
\begin{eqnarray*}
e^{\phi}& =& \sqrt{1-\frac{2M}{r}} \Rightarrow \phi = \frac{1}{2} \log \bigg(1- \frac{2M}{r}\bigg),\\
\Rightarrow \phi' &=& \frac{M/r^2}{1- \frac{2M}{r}} \Rightarrow \phi'' = \frac{(1- \frac{2M}{r}) (- \frac{2M}{r^3})-\frac{M}{r^2}\times\frac{2M}{r^2}}{(1- \frac{2M}{r})^2}\\
&=& \frac{- \frac{2M}{r^3}(1- \frac{2M}{r})}{(1- \frac{2M}{r})^2}.\\
b& =& \frac{r_{0}^2 + 2 M (r - r_{0})}{r} = 2M +\frac{r_{0}^2 - 2 M  r_{0}}{r},\\
b' &=& -\frac{r_{0}^2 - 2 M  r_{0}}{r^2}.\\
R^1_{010} &=& -\left(1-\frac{b}{r}\right)\bigg[\frac{- \frac{2M}{r^3}(1- \frac{2M}{r}) }{(1- \frac{2M}{r})^2} + \frac{\frac{M^2}{r^4}}{(1- \frac{2M}{r})^2} \bigg]- \frac{b' r -b}{2r^2} \phi'.
\end{eqnarray*}
At throat $ r = r_0$, first term is zero. Now
\begin{eqnarray*}
R^{\hat{1}'}_{\hat{0}'\hat{1}'\hat{0}'}|_{r = r_0}&=& - \frac{b' r -b}{2r^2} \phi'|_{r = r_0}\\
&=& \frac{1}{2 r^2_{0}}\bigg[\frac{r_{0}^2 - 2 M  r_{0}}{r_0} +\frac{r^2_0}{r_0} \bigg]\times\frac{\frac{M^2}{r^2_0}}{1- \frac{2M}{r_0}}\\
&=& \frac{1}{2 r^2_{0}}[2r_0-2M ]\times\frac{\frac{M^2}{r^2_0}}{1- \frac{2M}{r_0}}\\
&=& \frac{1}{ r^2_{0}}\frac{\frac{M}{r_0}(1- \frac{M}{r_0})}{1- \frac{2M}{r_0}}.
\end{eqnarray*}
Radial tidal acceleration of throat
\begin{eqnarray*}
\mid a^{\hat{i}'}\mid|_{r = r_0} = c^2 \xi' \times \frac{1}{ r^2_{0}}\frac{\frac{M}{r_0}(1- \frac{M}{r_0})}{1- \frac{2M}{r_0}}.
\end{eqnarray*}
For a traveler of size $\xi'= 2$ meter and $r_0 = 2.5 M = \frac{5}{2}M$,
\begin{eqnarray*}
\mid a^{\hat{i}'}\mid|_{r = r_0} = 2c^2  \times \frac{4}{25 M^2} \frac{\frac{2}{5}\times\frac{3}{5}}{\frac{1}{5}}= \frac{48 c^2}{125 M^2}.
\end{eqnarray*}
Now  $r_0 = 2.5 M $ is in $ G = c = 1$ unit. In dimension full unit $M = \frac{ G M }{c^2}$,
\begin{eqnarray*}
\mid a^{\hat{i}'}\mid|_{r = r_0} =  \frac{48}{125} \frac{c^2}{\frac{G^2 M^2}{c^4}} = \frac{48}{125}\times \frac{c^6}{G^2 M^2}.
\end{eqnarray*}
If we take $M$ to be of that of $ M87^\star$ as
\[
M= 6.5\times 10^{9} M_0 = 6.5 \times 10^{9}\times 2\times 10^{30}kg
 = 13\times 10^{39}kg, \] then\[
\mid a^{\hat{i}'}\mid|_{r = r_0} =  \frac{48}{125}\times\frac{(3\times 10^{8})^6}{(6.67\times 10^{-11})^2(13\times10^{39})^2}\]
\[\simeq~ 3.7\times10^{-10}m/s^2\]
\[\Rightarrow  \textrm{negligible because} ~~a'\sim \frac{1}{M^2}~~\textrm{and}~M ~\textrm{is large.}
\]
Similarly,  $a^2$ and $a^3$ will be small
$\Rightarrow$ Traversable.


\end{document}